\documentclass[12pt]{iopart}
\usepackage{xcolor}
\usepackage[numbers]{natbib}

\usepackage{url,microtype} 
\usepackage[onehalfspacing]{setspace}
\usepackage{dingbat}
\usepackage{longtable}
\usepackage{graphicx}
\usepackage{datetime}
\newdateformat{monthyeardate}{\monthname[\THEMONTH], \THEYEAR}
\usepackage{amsfonts}
\usepackage[textheight=8.5in]{geometry}
\usepackage{multirow}
\usepackage{comment}

\newcommand{\revision}{\color{black}}

\begin{document}

\title[Towards Human-Centric Intelligent Treatment Planning for Radiation Therapy]{Towards Human-Centric Intelligent Treatment Planning for Radiation Therapy} 

\author{Adnan Jafar, and Xun Jia*}

\address{Department of Radiation Oncology and Molecular Radiation Sciences, Johns Hopkins University, Baltimore, MD, USA}

\ead{xunjia@jhu.edu}
\vspace{10pt}
\begin{indented}
\item[]\monthyeardate\today
\end{indented}

\begin{abstract} 

Current radiation therapy treatment planning is limited by suboptimal plan quality, inefficiency, and high costs. This perspective paper explores the complexity of treatment planning and introduces Human-Centric Intelligent Treatment Planning (HCITP), an AI-driven framework under human oversight, which integrates clinical guidelines, automates plan generation, and enables direct interactions with {\revision operators}. {\revision We expect that} HCITP will enhance efficiency, potentially reducing planning time to minutes, and {\revision will} deliver personalized, high-quality plans. Challenges and potential solutions are discussed.

\end{abstract}


\maketitle

\section*{Introduction}

Cancer is the second leading cause of death globally, with 18.74 million new cases and 9.7 million cancer-related deaths reported in 2022 \cite{filhoglobocan}. Radiation Therapy (RT), which uses high-energy radiation to damage cancer cell DNA\cite{hall2006radiobiology}, is a cornerstone of cancer treatment, benefiting more than two-thirds of cancer patients, either as a standalone therapy or in combination with other modalities like surgery or chemotherapy. Modern RT techniques, such as intensity-modulated radiotherapy and volumetric-modulated arc therapy, enable precise control of a medical linear accelerator (LINAC) for radiation delivery that conforms to the tumor shape while sparing healthy tissues, resulting in reduced toxicity compared to conventional methods, as demonstrated in numerous clinical studies across diverse cancer types \cite{palma2008volumetric,gupta2012three,tribius2011intensity,nutting2011parotid}.

\begin{figure}[t]
    \centering
    \includegraphics[width=\linewidth]{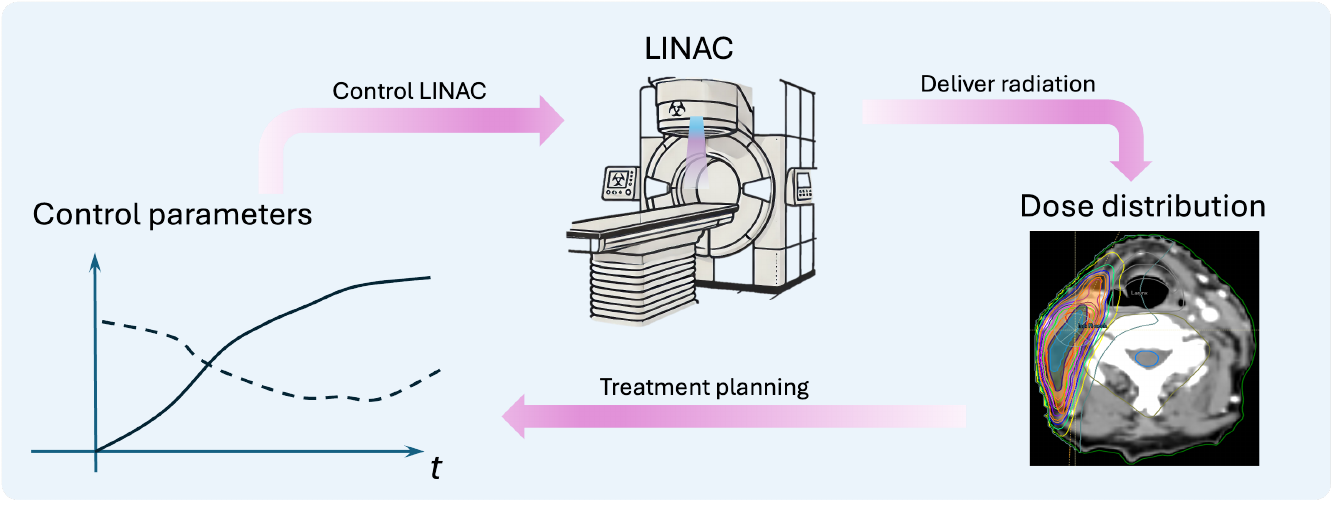}
    \caption{Relationship between control parameters, LINAC, and dose distribution in RT. A set of control parameters is input to the LINAC to control its motion and beam properties, generating a carefully sculpted dose distribution conformal to the target, while sparing doses to nearby organs. The treatment planning process refers to determining the LINAC control parameters for the patient-specific anatomy to yield a clinically acceptable dose distribution.}
    \label{fig:delivery}
\end{figure}

The success of RT critically depends on treatment planning, a foundational step determining the LINAC control parameters to specify its operations such as beam angle, radiation dose rate, and multi-leaf collimator motions to deliver the intended radiation dose (Figure~\ref{fig:delivery}) \cite{khan2014physics}. Plans must satisfy two criteria: \textit{deliverability}, which ensures physical execution by the LINAC, and \textit{acceptability}, which confirms alignment with treatment intent. Achieving these criteria currently relies on collaboration between planners and plan evaluators, e.g. physicians and medical physicists, using a Treatment Planning System (TPS), a specialized software that models radiation production and its interaction with patient-specific anatomy based on fundamental physics principles, and generates plans through mathematical optimization. Despite being the standard practice, this workflow suffers from suboptimal plan quality, low efficiency, and high costs, all of which negatively impact healthcare outcomes.

Historically, TPSs were designed to handle radiation physics modeling and plan optimization, delegating operational aspects to planners and allowing physicians to focus on patient care. However, a fundamental limitation of this workflow is the TPS's lack of intelligence, requiring extensive human input. In recent years, Artificial Intelligence (AI) has significantly transformed medicine, including RT. Advances have shown remarkable progress in decision making \cite{duan2019artificial}, outperforming humans in complex tasks \cite{silver2016mastering,fawzi2022discovering}. Building on these advancements, there is a growing opportunity to address the challenges in treatment planning. In this perspective article, we aim to shed some light on the complexities of the treatment planning process and potential solutions with AI-based decision-making capabilities. Such a solution has the potential to streamline the planning process, overcoming the limitations of the current practice and generating substantial impacts. 

\section*{Current Treatment Planning Practice and its Limitations}

Treatment planning begins with a physician defining a prescription, specifying the target dose for the tumor and tolerance doses for surrounding healthy tissues. The current practice then follows an iterative process involving two primary interactions (Figure~\ref{fig:workflow}). The first interaction is between a planner and TPS. After the planner defines dose distribution objectives in the objective function, the TPS solves the optimization problem while adhering to the LINAC’s physical constraints. The planner then repeatedly refines the objectives, guiding the TPS towards a plan that balances clinical objectives with technical feasibility. The second interaction involves the planner and the plan evaluators—typically the physician, who assesses the plan’s alignment with the clinical intent, and the medical physicist, who reviews its technical aspects. Feedback is then provided to the planner to further refine the plan. This cumbersome workflow presents significant limitations (Figure~\ref{fig:workflow}):

\begin{figure}[t]
    \centering
    \includegraphics[width=\linewidth]{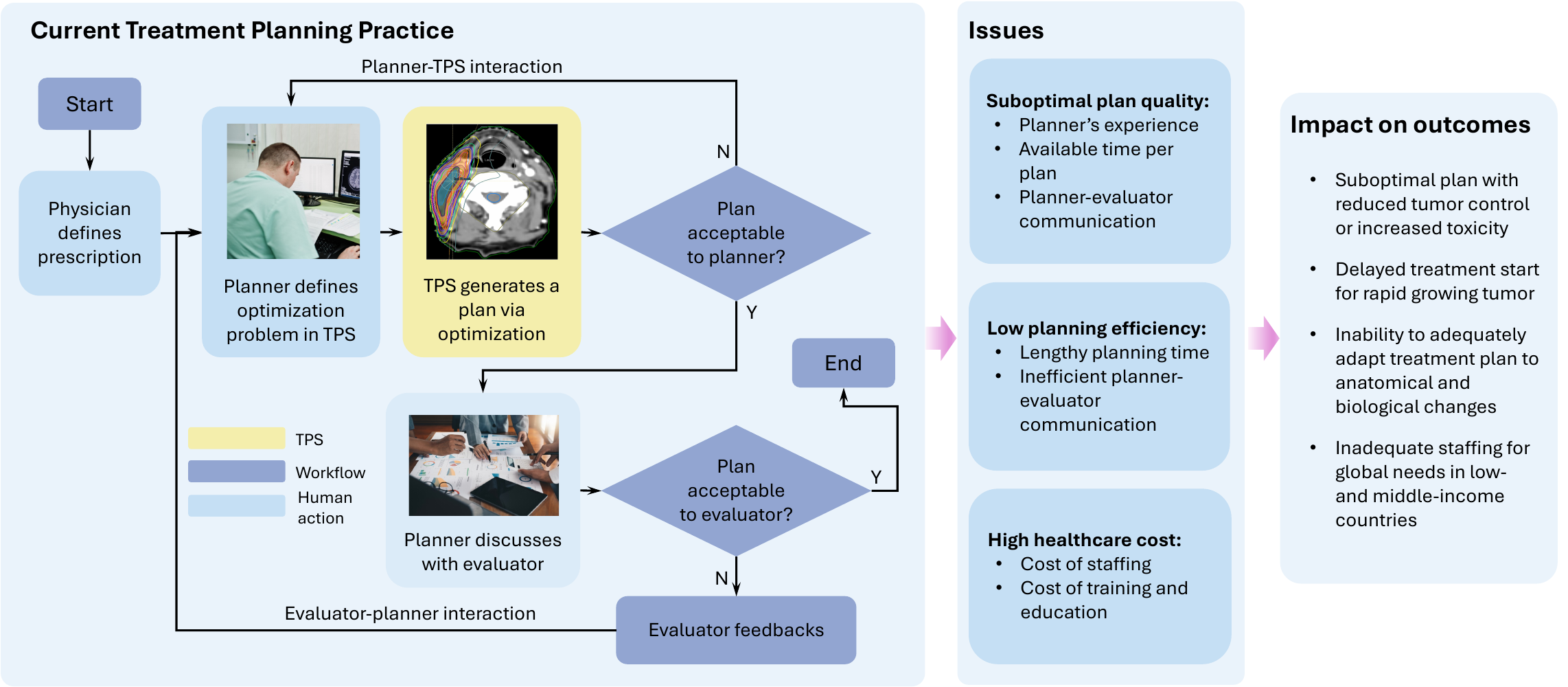}
    \caption{Current treatment planning workflow. After a physician defines a prescription, a planner repeatedly interacts with the TPS to define the objectives, which typically include adequate tumor coverage and maintaining normal tissue doses within tolerance levels. This interaction guides the TPS towards a solution that best meets these objectives while respecting the physical constraints of the LINAC, such as mechanical motion limits. The planner then discusses with plan evaluator about plan quality. This workflow causes several issues affecting treatment outcomes.}
    \label{fig:workflow}
\end{figure}

\textit{Suboptimal plan quality undermines treatment outcomes.} The optimal patient-specific plan is unknown, requiring planners to repeatedly interact with the TPS to explore the large solution space. The resulting plan quality heavily depends on human factors, including the planner's experience, planner-evaluator communication, and the time allocated for planning \cite{batumalai2013important, nelms2012variation}. Suboptimal plans, e.g. those with unnecessarily high dose to healthy tissues, are frequently accepted unknowingly \cite{das2008intensity}. An analysis of the RTOG-0126 clinical trial found that 9.1\% of patients received plans with unnecessarily 10\% higher normal tissue complication risks, which could have been avoided with better planning \cite{moore2015quantifying}. These plans deteriorate outcomes. In head-and-neck cancer, suboptimal plans have been associated with a 20\% lower 2-year overall survival and a 24\% higher 2-year local-regional failure rates \cite{peters2010critical}.

\textit{Low planning efficiency delays treatment and impacts outcomes.} The trial-and-error interaction between the planner and the TPS requires hours to generate a plan, while additional evaluator-planner iterations can extend this process to days or longer for complex cases. This tedious process prolongs the interval between diagnosis and RT initiation, which significantly impedes treatment outcomes. For example, in high-grade gliomas, each day of delay increases the risk of death by 2\% \cite{do2000effect}, and in head and neck cancer, RT delays can reduce loco-regional control by up to 12–14\% per week \cite{ferreira2015effect}. Additionally, delayed planning increases the chance of anatomical changes during the waiting period, making the plans for the initial anatomy suboptimal at the time of delivery, while also exacerbating patient anxiety and distress. Notably, with the rising global incidence of cancer \cite{soerjomataram2021planning}, a 15\% increase in new cases could lead to a 22.5\% rise in waiting times \cite{babashov2017reducing}, highlighting the urgent need to streamline RT planning processes and mitigate treatment delays.

\textit{High costs burden healthcare systems.} The current planning paradigm requires hospitals to hire professional planners, with a minimum ratio of one per 250 patients annually \cite{american2012safety}. This translates into significant costs for training (~\$145k per person) \cite{van2017cost}, salaries (median ~\$140k per person in US in 2023), and other expenses that are ultimately passed on to patients and healthcare systems.

These limitations are particularly pronounced in adaptive RT \cite{yan1997adaptive,li2013automatic}, which frequently adjusts treatment plans to account for anatomical changes. Replanning tasks demand stringent plan quality under tight time constraints. In online adaptive RT, where planning occurs while the patient is on the treatment couch, planning must be completed within minutes—a daunting task under the current practice. The limitations are further amplified in low- and middle-income countries, where more than 50\% of cancer patients requiring RT lack access to treatment \cite{zubizarreta2015need}. While efforts have been made to establish basic RT infrastructure like LINACs, the scarcity of trained personnel for treatment planning remains a critical bottleneck.

\section*{Existing Efforts using AI to Advance Treatment Planning}

Substantial efforts have been made to address these limitations over the years. For example, knowledge-based planning builds predictive models to derive patient-specific optimal dose-volume histograms (DVHs), a widely used measure representing the radiation dose distributions within specific structures for evaluating plan quality, to guide treatment planning \cite{ge2019knowledge}. In recent years, studies have incorporated AI technologies in this area. Our literature review (workflow in Supplementary Figure 1) identified existing studies, which can be broadly categorized into two groups.

The first group included studies focusing on the \textit{acceptability} criterion (Supplementary Table 1). Most studies leveraged deep neural networks to predict optimal dose distributions tailored to a patient’s anatomy \cite{nguyen2019feasibility, chen2019feasibility}. Yet, a key challenge remained—the deliverability of the predicted dose. As a result, these predictions primarily served as guidance for planners, who must use the TPS to approximate the predicted dose. This group also included studies that developed metrics to assess plan acceptability, providing additional guidance during treatment planning \cite{Gao2022Modeling}.

The second group of studies emphasized the \textit{deliverability} criterion (Supplementary Table 2). To replicate the decision-making process of human planners, researchers employed reinforcement learning (RL) and other techniques to develop virtual planners capable of operating the TPS \cite{shen2019intelligent, shen2021improving, wang2023integrated, gao2024multiagent, yang2024understanding}. These virtual planners have demonstrated performance comparable to human planners in head-to-head treatment planning competitions organized by scientific societies \cite{gao2024human, gao2023implementation}. More recently, Large Language Models (LLMs) were explored for autonomously adjusting organ priority weights \cite{liu2024automated}. Additionally, studies attempted to directly predict LINAC control parameters based on patient anatomy \cite{hrinivich2020artificial} {\revision using a deep Q-network method \citep{mnih2015human}}.

Based on the literature review, existing attempts have focused on addressing the two key criteria —\textit{deliverability} and \textit{acceptability}—separately. Moreover, a critical gap remains: these AI-based tools lack mechanisms for seamless interaction with physicians to incorporate their feedback, which is essential, as physicians are ultimately responsible for plan approval. With recent advances in AI demonstrating remarkable progress in decision-making and human-AI interaction \cite{duan2019artificial}, it is both timely and feasible to rebuild the treatment planning paradigm. 

\section*{Human-Centric Intelligent Treatment Planning}

\subsection*{Overall Scheme}
 
\begin{figure}[b]
    \centering
    \includegraphics[width=1.0\linewidth]{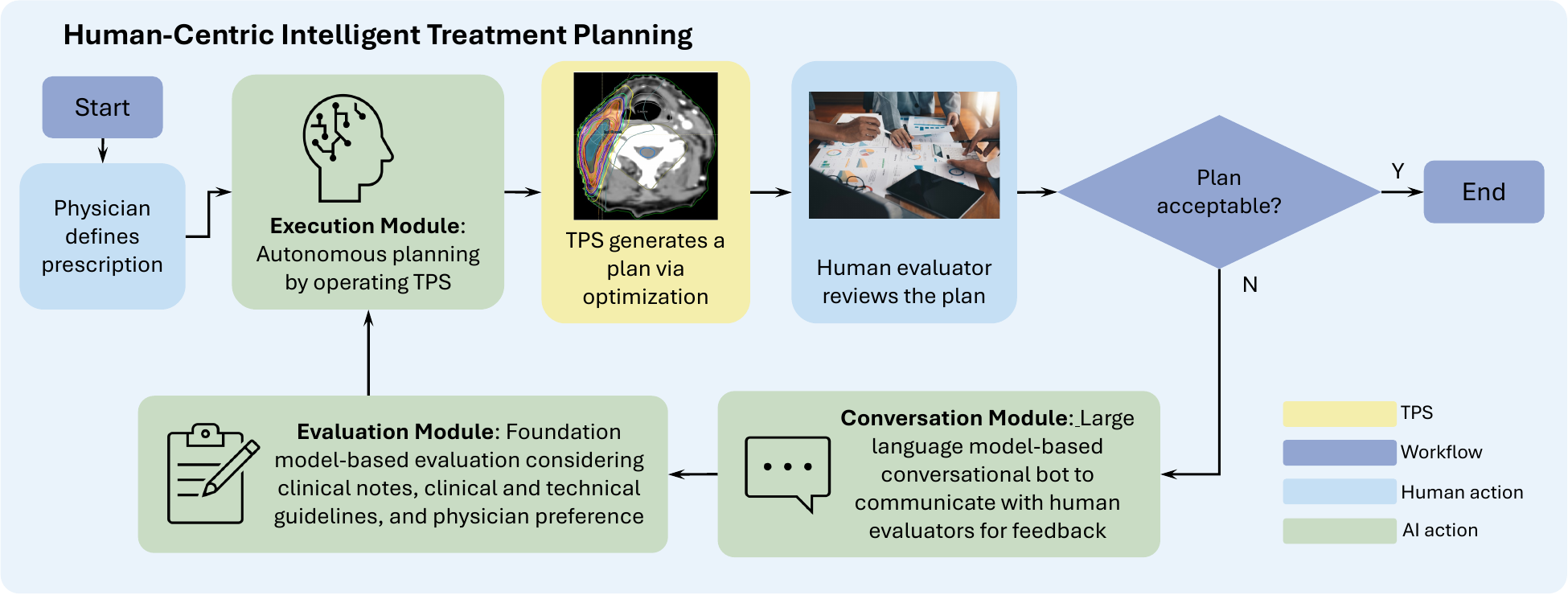}
    \caption{New treatment planning workflow enabled by HCITP. Under the guidance of the Evaluation Module, the Execution Module autonomously operates the TPS to generate a plan. {\revision Human evaluator reviews the plan} and provides feedback though the Conversation Module.}
    \label{fig:newworkflow}
\end{figure}

We envision the next-generation treatment planning workflow (Figure~\ref{fig:newworkflow}), Human-Centric Intelligent Treatment Planning (HCITP), enabled by a virtual planner {\revision composed of three} decision-making modules {\revision (highlighted in green)} to augment the TPS and interact directly with the human evaluator. Specifically, once the physician’s prescription is completed, HCITP immediately generates a preliminary treatment plan for review. The {\revision human} evaluators, typically the physician focusing on clinical aspects and the medical physicist addressing technical considerations, provide feedback to refine the plan. This feedback is communicated directly to the virtual planner for implementation. The resulting iterative loop between the evaluators and the virtual planner facilitates rapid completion of the planning workflow while maintaining high plan quality. {\revision Notably, HCITP serves as a tool to facilitate treatment planning, with the final responsibility for plan approval and conflict resolution always resting with the physician to ensure that clinical priorities and patient-specific considerations are upheld.}


At a high level, HCITP comprises three purposefully designed core components. The first is the \textit{Evaluation Module}, which is responsible for assessing the quality of the treatment plan. Built on foundation models (FMs) with explainable AI techniques \cite{holzinger2022explainable}, it evaluates the plan with respect to clinical guidelines, physician's preference, as well as practical considerations. FMs refer to large-scale machine learning models trained on broad and diverse data that can be adapted to a wide range of downstream tasks with minimal task-specific tuning, serving as a versatile backbone for many AI applications.\cite{yi2024open, zhou2024comprehensive} With FM, the Evaluation Module processes multimodal data, including clinical protocols, technical guidelines, medical images of various modalities, treatment plans, clinical notes, etc., to generate contextualized embedded states for evidence-driven, {\revision case-adapted} treatment plan assessment. In addition, the module incorporates physician preferences, such as trade-offs among organ doses. While a plan may meet standard quality guidelines, it can still be rejected due to individual physician preferences. To address this, the Evaluation Module encodes these preferences based on historically approved plans, enabling HCITP to prioritize the plans most likely to receive physician approval. Furthermore, this module also assesses practical aspects of the plan related to deliverability, such as delivery time and plan modulation factor {\revision that reflects the complexity of a plan and hence the level of accuracy required by the LINAC to precisely deliver it}. When building this module, explainable AI can be employed to make the decision-making processes transparent and understandable to humans \cite{holzinger2022explainable}. We also envision that a key feature of this module is continual learning, allowing it to monitor and integrate the up-to-date clinical and technical guidelines, ensuring evaluations remain aligned with current standards. {\revision The FM-based Evaluation Module can be built on task-specific encoders, for example language models such as BioBERT \citep{lee2020biobert} and PubMedBERT \citep{gu2021pubmedbert} for clinical guidelines, imaging models such as nnU-Net \citep{isensee2018nnunet} and Swin-UNETR \citep{hatamizadeh2021swinunetr} for medical images, and NLP tools such as cTAKES \citep{savova2010ctakes} and ClinicalBERT \citep{huang2019clinicalbert} for clinical notes. Multimodal integration can be achieved through fusion at low-level features (early fusion), at encoder level (mid fusion), or at the decision stage (late fusion), as well as through hybrid approaches, enabling the FM to reason jointly across modalities and to support case-adapted plan assessment for the execution module}.

Second, the \textit{Execution Module} replicates decision-making capability of human planners and autonomously operates the TPS, aiming at generating deliverable treatment plans under the guidance of the Evaluation Module. This module can be built using RL, a machine learning technique in which an agent learns to make decisions through interactions with an environment. The goal is to discover policies that achieve specific objectives, such as treatment planning in HCITP, by maximizing a reward function, which serves as a numerical signal that reflects how favorable each decision is with respect to the defined goals \cite{sutton2018reinforcement}. The reward function is derived from the Evaluation Module to quantify how well the treatment plan satisfies both clinical criteria and practical considerations. Training this module should incorporate human experience in operating the TPS. To enhance versatility, FMs may be used as the underlying architecture. The training dataset should include cases with diverse tumor sites, patient anatomies, and clinical conditions. 


The third one is the \textit{Conversation Module} powered by LLMs and speech recognition technologies. Its purpose is to keep humans in the loop under a smooth workflow by enabling interactive feedback and guidance throughout the planning process—hence the term human-centered in HCITP. This module enables real-time bi-directional communication with the evaluator by summarizing feedback on plan quality and prompting for clarification when needed. In {\revision contrast} to the current clinical workflow, where the human evaluator's feedback reaches the TPS indirectly through the human planner, the direct interaction between evaluators and the TPS ensures that clear, actionable input is relayed to the planning process, supporting real-time dynamic plan refinement.

\subsection*{Advantages of Human-Centric Intelligent Treatment Planning}


By establishing an AI-augmented treatment planning workflow under human's oversight, HCITP holds advantages over existing approaches and addresses several key challenges (Table~\ref{tbl:advantages}). 

\begin{table}
    \caption{Key Features of HCITP and its advantages compared to the current treatment planning workflow.}
    \centering
    \footnotesize
    \begin{tabular}{p{2.2cm}|p{6cm}|p{6cm}}
    \hline
       Key features  & Advantages & Existing approaches and challenges \\
    \hline
       Evaluation Module & Plan quality evaluation incorporating the latest clinical and technical guidelines and physician preferences, prioritizing plans most likely to receive approval. & Evaluations based on institutional guidelines, which may not be up-to-date. Physician preferences incorporated through evaluator-planner interaction. \\
       \hline
    Execution Module & RL exploration for plan quality not limited by training data. High-quality plans and planning strategies to facilitate education. & Human planner operating TPS with limited exploration of plan quality. \\
    \hline
    Conversation Module & Natural and intuitive way to intake evaluator’s feedback for streamlined workflow. & Evaluator instructs planner to implement changes, impeding workflow due to the extra layer of communication. \\
    \hline
    Human evaluator in the loop & Plan approval from decision maker. & Evaluator in the loop, but relying on planner to implement changes. \\
    \hline
    TPS in the loop & Plan deliverability ensured via physics principles and LINAC modeling. & TPS in the loop, but relying on planner to operate it. \\
    \hline
    Continual learning & Automatic incorporation of the up-to-date clinical and technical guidelines. & Incorporation of the up-to-date clinical and technical guidelines by humans. \\
    \hline
    Scalability & Applications over RT clinics to reduce demand in human expertise in treatment planning. & Each clinic implements its own practice that heavily relies on human expertise in treatment planning. \\
    \hline
    \end{tabular}
    \label{tbl:advantages}
\end{table}

Leveraging the few-shot learning capabilities of FMs due to the extensive pre-training and their ability to process and contextualize multimodal data, HCITP is designed to manage across various cancer sites while continuously incorporating up-to-date plan evaluation criteria. Integrating physician preferences with clinical guidelines ensures that plans are optimized not only for clinical quality but also for individual patient needs and physician-specific standards, enhancing personalization.

In terms of generating plans to meet treatment intent, the exploratory nature of RL enables the Execution Module to uncover novel planning strategies, potentially pushing the boundaries of achievable plan quality beyond existing clinical practices. This also provides educational value by offering insights into optimal treatment plans and planning strategies. The deliverability criterion is maintained through the direct integration of the Execution Module with the TPS, ensuring adherence to physics principles, machine constraints, and other practical requirements.

HCITP also streamlines the workflow by allowing human evaluators to provide direct, natural, and intuitive feedback to the planning process, which is dynamically processed by the Evaluation Module and then passed on to the Execution Module. This maintains human oversight for the treatment planning process, and eliminates the intensive and iterative task of planners manually interpreting and encoding evaluators' feedback in the current clinical workflow, holding the potential to reduce planning time from days to minutes. The resulting reduction in planning time will shorten the interval between diagnosis and treatment initiation, critical for improving outcomes, particularly in rapidly progressing tumors.

Finally, by reducing reliance on human planners, HCITP has the potential to lower costs and expand access to RT services, especially in resource-limited settings, ultimately enhancing global cancer care.

Notably, previous studies have explored similar concepts of human-centric RT planning, albeit under different terminologies \cite{callens2024full, sheng2021artificial, estro2023ai}. A particularly relevant analogy has been drawn between aviation and RT. In aviation, automation has been seamlessly integrated under pilot oversight, shifting the pilot’s role from direct control to system management while maintaining their critical decision-making authority \cite{callens2024full}. Similarly, in RT, automation is expected to enhance treatment planning without diminishing the essential roles of human experts. However, a key distinction in the current RT planning workflow lies in the division of human roles: planners generate treatment plans, while physicians approve them and provide feedback. This introduces an added layer of complexity: planner-physician communication, unlike the pilot model in aviation. To address this, HCITP redefines the workflow by positioning the physician as the central human component, directly interacting with AI automation through the Conversation Module, thereby streamlining the process and reducing inefficiencies.

\section*{Considerations on Human-Centric Intelligent Treatment Planning}

Given the revolutionary nature of HCITP, there are foreseeable challenges that call for our prompt actions towards the effective development of this system. 

\subsection*{Challenges Related to Technology Development}
\

\textit{Model Training:} Well-validated, trusted data form the foundation for training the HCITP model \cite{chen2018why, esteva2017dermatologist}. As with developing any AI-driven systems, collecting and curating such data presents significant challenges. Training the Evaluation Module should include clinical and practical guidelines on plan evaluation. Because it also assesses plans in the context of physician preferences, data collection efforts should include gathering physician-specific prior multimodal treatment plans paired with corresponding physician decisions. Planners' actions in operating the TPS in the current practice and conversation data between physicians and human planners may be collected to train the Execution Module and the Conversation Module. 
Meanwhile, powerful generative models, like diffusion models, may be employed to synthesize data, reducing the burden for extensive data acquisition. Yet expert review by RT professionals is necessary to verify the plausibility and clinical relevance of the data. As the HCITP modules are developed, the integration of explainable AI techniques is essential to ensure transparency, trustworthiness, and reliability in the clinical decision-making process for RT treatment planning \cite{thirunavukarasu2023large, wu2024usable}.

From the computational standpoint, training the Execution Module via the RL framework requires repeated interactions with the TPS to learn an optimal policy for operating it. This is computationally intensive, as the solution space expands rapidly when exploring complex operation strategies that experienced human planners can master. The challenge becomes even more significant for anatomically complex cancer sites . To mitigate these issues, enhancing the computational power of the TPS to accelerate the solution of plan optimization problems is essential. Augmenting the RL training process with human experience in planning decisions can guide the RL agent’s exploration and facilitate faster convergence \cite{grondman2012survey}.

\textit{Variability in Acceptability and Deliverability:} Treatment plans in current clinical practice often exhibit substantial variability in both acceptability (e.g., plan quality and clinical trade-offs) and deliverability (e.g., machine limitations and treatment complexity). This poses a challenge for training HCITP, as it introduces ambiguity in defining what constitutes an optimal plan. The variation in acceptability is multifaceted. One major factor is the lack of a definitive ground truth for plan quality evaluation. With HCITP, state-of-the-art clinical guidelines can be integrated. Additionally, HCITP will learn physician preferences for plan acceptance. The Evaluation Module, trained in this way, will provide guidance to the Execution Module, ensuring greater consistency in generated plans. Another key factor contributing to variability is the acceptance of suboptimal plans due to time constraints or ineffective communication between planners and physicians. HCITP’s streamlined workflow facilitates the pursuit of optimal plans, thereby reducing quality variations. Moreover, by incorporating clinical guidelines and enabling physicians to explore a broader range of plans, HCITP can offer valuable educational opportunities, helping to mitigate variations driven by individual human factors. Regarding variability in deliverability, the Evaluation Module will be trained not only to assess plan quality from a clinical perspective, but also to account for other practical factors, such as plan modulation factors, delivery time considering patient tolerance, beam angles appropriate for immobilization devices to prevent collisions, and more. Recognizing this variability, HCITP development will likely need to be iterative. Early, controlled implementation can standardize planning strategies, reduce unwarranted variation, and supporting the system’s continual refinement.

\textit{Generalization:} While ensuring generalization across datasets for diverse populations is critical, in treatment planning, generalization also refers to the ability to perform this task for a wide range of tumor sites. Unlike human planners, who are trained to handle various disease sites, existing virtual planners are developed for specific cancer types, limiting their scalability and versatility. RL-based Execution Module can be effectively trained to incorporate broad knowledge in operating the TPS and be fine-tuned for different tumor sites.

{\revision HCITP leverages published clinical guidelines, e.g. those from the American Society for Radiation Oncology and the European Society for Radiotherapy and Oncology etc. as part of its FM pre-training to ensure broad generalizability. However, alignment with local datasets and institutional protocols should not be neglected for safe and clinically relevant deployment. This alignment can be achieved through strategies such as fine-tuning on de-identified local data, federated learning across institutions, or feedback loops that allow the model to continuously adapt to local practice patterns.}

\textit{Continual Learning:} RT and treatment planning continuously evolve to accommodate advancements such as new treatment guidelines and innovative delivery approaches of LINACs. To keep pace with them, HCITP should be designed to seamlessly monitor and integrate with society guidelines and diverse treatment delivery technologies, facilitating the adoption of the latest RT techniques and protocols. Transfer learning can be employed to reduce the effort required for training and implementation. Additionally, regular audits on data quality are necessary to detect and address emerging biases or performance deficiencies. A robust feedback loop should be incorporated, allowing users to provide input during routine clinical practice to refine and enhance the system’s performance.

\subsection*{Challenges Related to Clinical Implementation}
\

\textit{Model Development and Deployment:} Implementing HCITP requires significant investment, such as computational and data resources for training, as well as infrastructure to support model inference at deployment. This may not be feasible universally across hospitals, especially in resource-constrained settings globally. To address this challenge, we envision using  lightweight models that can run locally on multiple GPUs, with the option to leverage cloud resources when necessary. Rather than full model training, a more practical approach involves using lightweight post-training techniques. 

It is important to acknowledge that HCITP performance may not always be perfect. It is intentionally designed to maintain human oversight, with physicians being the ultimate decision-makers in approving treatment plans. The AI modules support key tasks in the workflow, relieving human planners from repetitive and routine duties. However, human planners are expected to continue to play a critical role in managing complex cases beyond AI's current capabilities. 

Meanwhile, overemphasizing human-centeredness may inadvertently limit plan quality improvement and educational opportunities. To prevent this, the Evaluation Module should prioritize that ensuring the latest planning guidelines are followed in generated plans, promoting consistency and improving plan quality across institutions. Additionally, the enhanced workflow efficiency by HCITP will allow physicians more time to thoroughly review and refine plans. By observing a broader range of plans and exploring the solution space, physicians can better identify the optimal plans for individual patients. This process also provides valuable educational opportunities.

\textit{Evaluation:} High-quality representative datasets must be collected and infrastructure must be built to support evaluation. A well-defined pathway of evaluation should be established, starting with offline virtual testing on large-scale independent datasets, followed by pilot studies. {\revision Rigorous uncertainty estimation, e.g. via ensembles and Monte Carlo dropout\citep{he2023survey}, calibration, and robustness testing should be performed.} Ultimately, a prospective evaluation, akin to multi-center clinical trials, should be conducted to objectively measure the overall impact of HCITP on patient care and healthcare delivery. Post-deployment, regular audits with diverse clinical data are necessary to monitor and sustain safety and performance. 

While technical metrics, such as cumulative rewards and convergence rates, can provide insights into the performance of AI models, it is more important to design task-based metrics to assess HCITP's performance in a contextualized setting. For example, plan quality can be measured using established numerical models to measure impact on healthcare outcomes \cite{bentzen2010quantitative}. Health-economics models may be employed to evaluate the cost-effectiveness of HCITP implementation \cite{lievens2012health}. For explainability, HCITP’s strategies in generating plans and evaluating them can be compared against those of expert humans to validate their effectiveness and alignment with clinical expertise.


\textit{Safety and Privacy:} FMs can sometimes lead to hallucinations or incorrect outputs \cite{xu2024hallucination}, posing risks to patient safety. Risks may also arise from improper explorations in RL model training, poorly designed reward functions, and biased training data, all of which can result in suboptimal or discriminatory actions. Differences in LINAC and TPS functions, compatibility, and dose modeling accuracy may introduce systematic biases during model training. Additionally, adversarial attacks on internal training data could lead to harmful or misleading outputs. 

As for privacy, large-scale AI systems, particularly FMs, are often trained on vast datasets that may contain personal information, raising concerns about data privacy. Malicious actors could exploit vulnerabilities, such as prompting tricks, to manipulate the models into revealing sensitive protected health information, thereby violating confidentiality standards. Such breaches could lead to legal consequences, erosion of trust in healthcare technologies, and increased patient reluctance to consent to AI-assisted care.

Several strategies can help mitigate these issues. For instance, combining chain-of-thought prompting, which guides the model to reason step-by-step, with self-consistency, which improves reliability by generating multiple reasoning paths and selecting the most frequent or confident response, has been shown to enhance LLM reasoning accuracy by 5–10\% \cite{huang2022large, wang2022self}. Retrieval-augmented generation can further improve model responses by incorporating relevant external information. Additionally, guardrails such as regular model evaluations, adversarial testing, and continuous monitoring post-deployment are essential. 

\textit{Legal Considerations and Clinical Adoption:} While potentially revolutionizing RT treatment planning, HCITP also raises a critical question common to AI-based healthcare systems: who should be held accountable for errors it makes? In the HCITP framework, much like the current practice, physicians retain the authority to approve or reject plans. This ensures that physicians remain ultimately accountable for their validity.

As with the adoption of other AI techniques in healthcare, government guidelines are essential to establish clear roles and responsibilities for all parties involved. Software manufacturers must prioritize creating reliable AI systems, rigorously testing them across diverse datasets to ensure robustness, and transparently disclosing system limitations. Users, in turn, should undergo comprehensive training to effectively interpret AI-generated recommendations and validate their applicability before implementation. During operation, identified errors should be reported. By fostering collaboration between users and software manufacturers, and strengthening these efforts through robust legislation, the risks associated with HCITP can be minimized. {\revision To obtain regulatory approval, vendors developing HCITP systems must demonstrate compliance with medical-device standards (e.g., U.S. FDA requirements), providing validated evidence of accuracy, transparency, reproducibility, and robustness in treatment planning, while also addressing ethical and privacy concerns.}

{\revision It is essential to strategize a roadmap that builds trust among key stakeholders including patients, clinicians, administrators, regulators, and vendors. This roadmap should prioritize focused technology development on the key attributes outlined above, be supported by comprehensive multi-site validation that benchmarks performance against expert practice, and ensure alignment with regulatory and ethical standards. Throughout development and deployment, continuous human oversight with clearly defined responsibilities should be maintained. Following pilots and controlled trials, a phased roll-out with structured user training should be implemented, and objective evidence of effectiveness should be reported routinely to guide scale-up and continuous improvement. Early, sustained stakeholder engagement will help streamline approval, foster confidence, and increase the likelihood of successful adoption.}

\section*{Conclusion}

This perspective paper outlines key challenges in current RT treatment planning, particularly the lack of intelligence within existing TPSs. {\revision As a solution, we envision HCITP as a unified, AI-powered framework that integrates decision-making capabilities while preserving human oversight to ensure quality and safety. Unlike prior efforts that address isolated aspects of treatment planning, HCITP aims to harmonize these solutions into a single workflow.} We look forward to future developments in this area, highlighting the potential for HCITP to enhance personalized treatment planning, increase access to RT, and drive significant improvements in clinical practice.

\section*{Acknowledgements}

This work was supported in part by NIH grants R01CA227289, R01CA254377, R37CA214639, and R01EB032716. 

\section*{Author Contributions}

A.J. and X.J. wrote the main manuscript text and prepared figures. Both authors have read and approved the manuscript.

\section*{Competing Interests}

The authors declare no competing interests.

\bibliographystyle{unsrtnat}
\bibliography{reference}

\newpage
\section*{Figure Legends}
\noindent Figure~\ref{fig:delivery}. Relationship between control parameters, LINAC, and dose distribution in RT. A set of control parameters is input to the LINAC to control its motion and beam properties, generating a carefully sculpted dose distribution conformal to the target, while sparing doses to nearby organs. The treatment planning process refers to determining the LINAC control parameters for the patient-specific anatomy to yield a clinically acceptable dose distribution.

\

\noindent Figure~\ref{fig:workflow}. Current treatment planning workflow. After a physician defines a prescription, a planner repeatedly interacts with the TPS to define the objectives, which typically include adequate tumor coverage and maintaining normal tissue doses within tolerance levels. This interaction guides the TPS towards a solution that best meets these objectives while respecting the physical constraints of the LINAC, such as mechanical motion limits. The planner then discusses with plan evaluator about plan quality. This workflow causes several issues affecting treatment outcomes.

\

\noindent Figure~\ref{fig:newworkflow}. New treatment planning workflow enabled by HCITP. Under the guidance of the Evaluation Module, the Execution Module autonomously operates the TPS to generate a plan. Human evaluator reviews the plan and provides feedback though the Conversation Module.

\section*{Table Legend}
\noindent Table~\ref{tbl:advantages}. Key Features of HCITP and its advantages compared to the current treatment planning workflow.

\end{document}